# Magnetism and superconductivity in $Ru_{1-x}Sr_2RECu_{2+x}O_{8-d}$ (RE=Gd, Eu) and $RuSr_2Gd_{1-y}Ce_yCu_2O_8$ compounds

P.W. Klamut [1], B. Dabrowski [1], S. M. Mini [1], S. Kolesnik [1], M. Maxwell [1], J. Mais [1],
A. Shengelaya [2], R. Khazanov [2,3], I. Savic [2,4], H. Keller [2], C.Sulkowski [5], D.Wlosewicz [5], M. Matusiak [5],
A.Wisniewski [6], R.Puzniak [6], I.Fita [6]

[1] Department of Physics, Northern Illinois University, DeKalb, Illinois 60115, USA
[2] Physik-Institut der Universität Zürich, CH-8057 Zürich, Switzerland
[3] Laboratory for Muon-Spin Spectroscopy, Paul Sherrer Institut, CH-5232 Villigen PSI, Switzerland
[4] Faculty of Physics, University of Belgrade, 11001 Belgrade, Yugoslavia
[5] Institute of Low Temperature and Structure Research of Polish Academy of Sciences, 50-950 Wroclaw, Poland
[6] Institute of Physics of Polish Academy of Sciences, 02-668 Warszawa, Poland

We discuss the properties of new superconducting compositions of $Ru_{1-x}Sr_2RECu_{2+x}O_{8-d}$ (RE=Gd, Eu) ruthenocuprates that were synthesized at 600 atm. of oxygen at 1080°C. By changing ratio between the Ru and Cu, the temperature of superconducting transition ($T_C$) raises up to $T_C^{max}$=72 K for x=0.3, 0.4. The hole doping achieved along the series increases with Cu→Ru substitution. For $x\neq0$, $T_C$ can be subsequently tuned between $T_C^{max}$ and 0 K by changing oxygen content in the compounds. The magnetic characteristics of the RE=Gd and Eu based compounds are interpreted as indicative of constrained dimensionality of the superconducting phase. Muon spin rotation experiments reveal the presence of the magnetic transitions at low temperatures ($T_m$=14-2 K for x=0.1-0.4) that can originate in the response of Ru/Cu sublattices. $RuSr_2Gd_{1-y}Ce_{1-y}Cu_2O_8$ ($0\leq y\leq0.1$) compounds show the simultaneous increase of $T_N$ and decrease of $T_C$ with y. The effect should be explained by the electron doping that occurs with Ce→Gd substitution. Properties of these two series allow us to propose phase diagram for 1212-type ruthenocuprates that links their properties to the hole doping achieved in the systems. Non-superconducting single-phase $RuSr_2GdCu_2O_8$ and $RuSr_2EuCu_2O_8$ are reported and discussed in the context of the properties of substituted compounds.

Recent reports of the apparent coexistence of superconductivity (SC) and ferromagnetism (FM) in ruthenocuprates [1,2] have triggered intense interest in the properties of these materials. The compounds that exhibit this unusual behavior are $RuSr_2RECu_2O_8$ (Ru-1212) [3] and $RuSr_2(RE_{2-x}Ce_x)Cu_2O_{10-y}$ (Ru-1222) (RE=Gd, Eu) [2] and they belong to the family of high temperature superconductors (HTSC). Structurally similar to the well-known $GdBa_2Cu_3O_7$ (Gd123) superconductor, the ruthenocuprate $RuSr_2GdCu_2O_8$ is a layered perovskite containing both $CuO_2$ and $RuO_2$ planes in its crystal structure. The correspondence between the two structures can be described by replacing the so called chain -Cu atoms in the Gd123 by Ru ions coordinated with full octhaedra of oxygens thus forming $RuO_2$ planes. The positions of Sr atoms in Ru-1212 correspond to Ba positions in Gd123, and the structural block containing $CuO_2$ double planes remains similar for both compounds. In $RuSr_2GdCu_2O_8$, the magnetically ordered state manifests itself at temperatures $T_N$=130-136 K, much higher than the superconducting transition reported at 45 K for the highest $T_C$ samples. The magnetic order persists in the superconducting state [2]. What makes these compounds unique in the family of HTSC is that the magnetic ordering originates in the sublattice of the d-electron Ru ions. Recent muon spin rotation and magnetization results provided evidence for the coexistence of the magnetic ordering of Ru moments with superconductivity at low temperatures [2]. Although the ferromagnetic ordering was initially proposed for the Ru sublattice below $T_N$ [2], recent neutron diffraction experiments show that the dominant magnetic interactions are of the G-type antiferromagnetic (AFM) structure [4, 5]. Based on these results, the observed ferromagnetism should originate from the canting of the AFM lattice that gives a net moment perpendicular to the c-axis [5]. This scenario resembles the description of the properties of $Gd_2CuO_4$, a non-superconducting weak ferromagnet, where the distortions present in the $CuO_2$ plane permit the presence of antisymmetric superexchange interactions in the system of Cu magnetic moments [6, 7]. In the ruthenocuprates, however, the weak ferromagnetism originates in the $RuO_2$ planes and should result in the effective magnetic field being parallel to the $CuO_2$ planes [2, 5, 8].

## 1. $Ru_{1-x}Sr_2GdCu_{2+x}O_{8-d}$ ($0 \le x \le 0.75$)

In order to address how the properties of $RuSr_2RECu_2O_8$ can be affected by the dilution of the magnetic sublattice of Ru, we attempted to partially substitute Ru with Cu ions. With this substitution the nominal formula of the resulting compound should change toward the hypothetic $GdSr_2Cu_3O_7$ , a Sr containing analogue of the $GdBa_2Cu_3O_7$ $T_C \approx 92$ K superconductor. We have found that for $Ru_{1-x}Sr_2GdCu_{2+x}O_{8-d}$ series, the layered Ru-1212 type structure becomes stable only during synthesis at high pressure oxygen conditions. Polycrystalline samples of $Ru_{1-x}Sr_2GdCu_{2+x}O_{8-d}$ (x=0, 0.1, 0.2, 0.3, 0.4, 0.75) were prepared by solid-state reaction of stoichiometric $RuO_2$, $SrCO_3$, $Gd_2O_3$ and CuO. After calcination in air at 920°C the samples were ground, pressed into pellets, and annealed at 970°C in flowing oxygen. Then the samples were sintered at 1060°C for 10 hours in a high pressure oxygen atmosphere (600 bar). Repeated annealing at high pressure oxygen conditions improved the phase purity of the material while not changing their superconducting and magnetic properties. Fig. 1 presents changes of the lattice parameters with x. The insets to figure 1 show the x-ray diffraction patterns for the x=0.4 and 0.75 compositions.

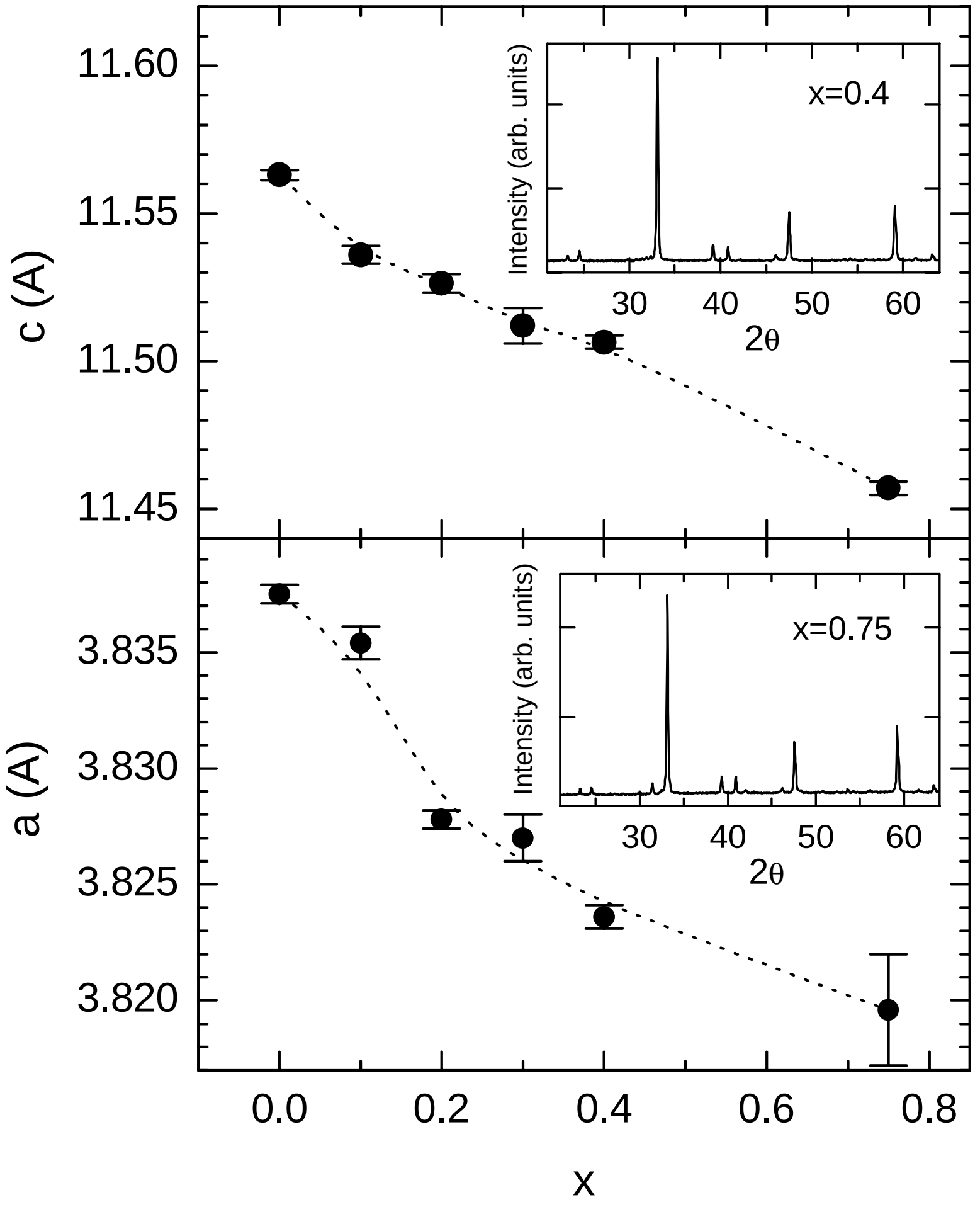

Fig. 1. Lattice constants for $Ru_{1-x}Sr_2GdCu_{2+x}O_{8-d}$. Insets show XRD patterns for x=0.4 and 0.75 samples

Both *a* and *c* dimensions decrease with the substitution of Cu for Ru in the Ru-O planes. The observed change can indicate increased hole doping with x for the series, which will be confirmed in further reported X-ray Absorption Near Edge Spectroscopy (XANES), thermopower, and Hall effect experiments. Figure 2 and 3 present the set of temperature dependencies of resistivity and *ac* susceptibility measured for this series. They are also compared to the same dependencies obtained for the parent x=0 compound, which was synthesized in flowing oxygen at 1060C followed by slow cooling.

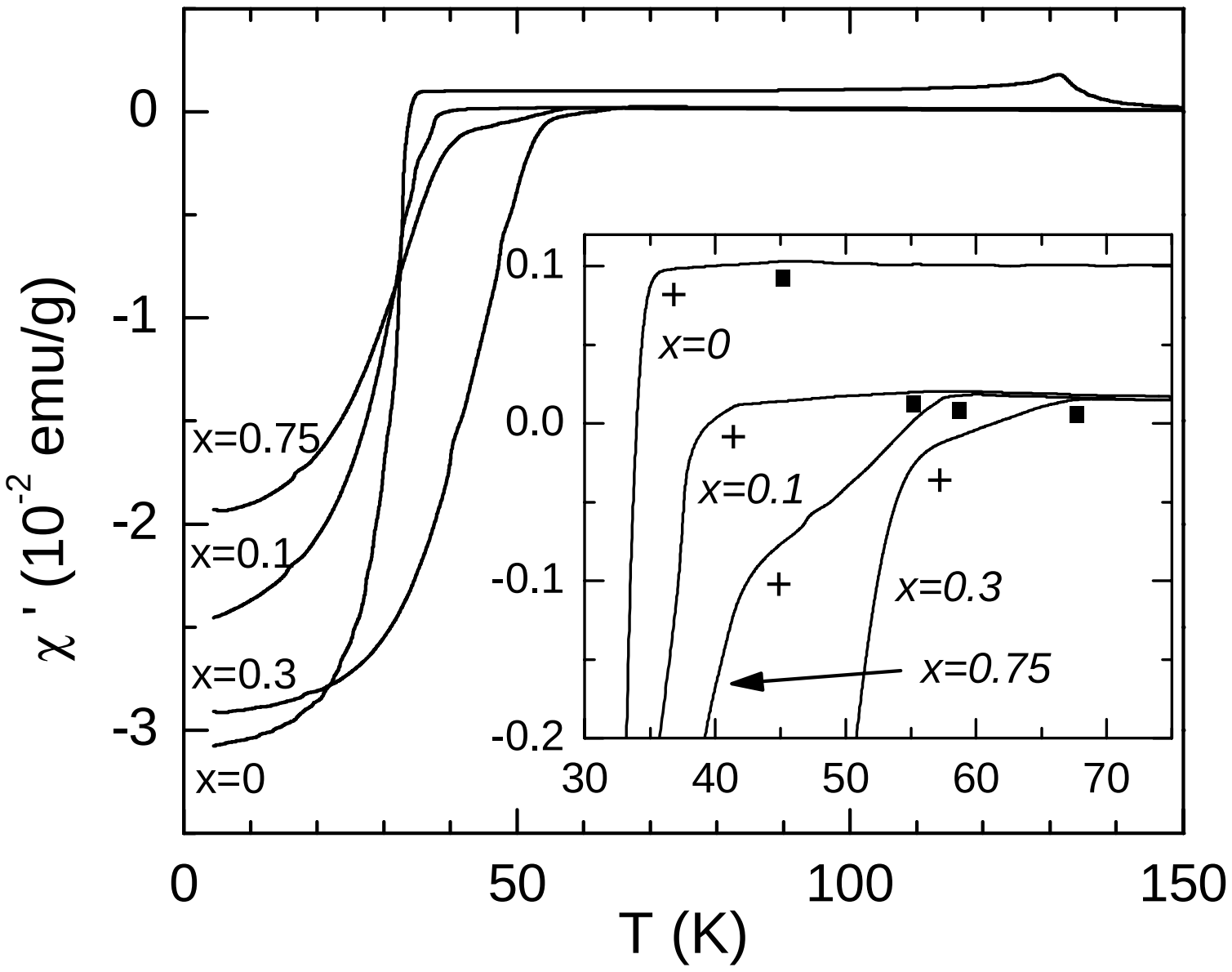

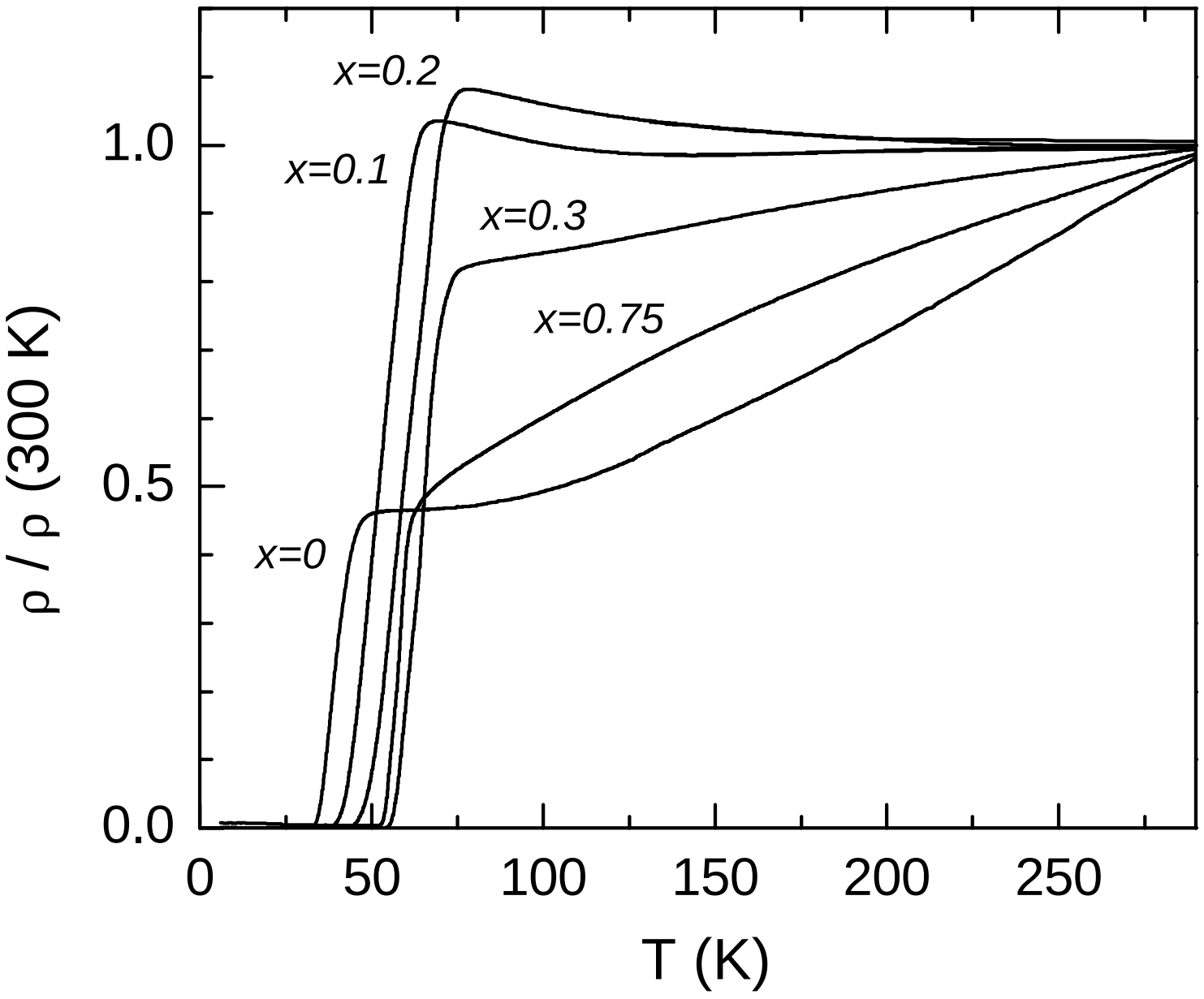

Figs. 2 and 3. The temperature dependencies *of ac* susceptibility ($H_{ac}$=1 Oe, f=200 Hz) and resistivity for $Ru_{1-x}Sr_2GdCu_{2+x}O_{8-d}$

The onset temperatures for the resistive superconducting transition increase from $T_C^{on}$ = 45 K for x=0 to 72 K for x=0.3 and 0.4, and then decrease to 62 K for the 0.75 sample. The *ac* susceptibility results indicate that the upturn at $T_N$=132 K, associated with the onset of the magnetic transition for the x=0 compound, is absent for all x≠0 samples. Fig. 4 presents the magnetic field dependencies of *dc* magnetization for the series measured at 4.5 K. The comparison of the M(H) dependence for $RuSr_2GdCu_2O_8$ with the behavior of non-superconducting $Gd^{3+}Ba_2Cu_3O_{6.2}$ (open circles in Fig.4, for the response of paramagnetic sublattice of $Gd^{3+}$ ions) reveals additional contribution to the magnetization observed in the parent ruthenocuprate. This may represent substantial ferromagnetic alignment of the Ru moments (approx. 1 $\mu_B$ per formula unit at $H_{dc}$=6 T) for the values of the magnetic field that are still well below the critical field for the superconducting phase [9]. This leads to the important conclusion that the weak ferromagnetism observed for superconducting Ru-1212 is significantly enhanced in the presence of the magnetic field. The x≠0 samples follow practically the same dependence originating in the response of paramagnetic ions in the $Gd^{3+}$ sublattice. This shows that the type of magnetic order characteristic of the parent ruthenocuprate is absent in the Cu rich samples.

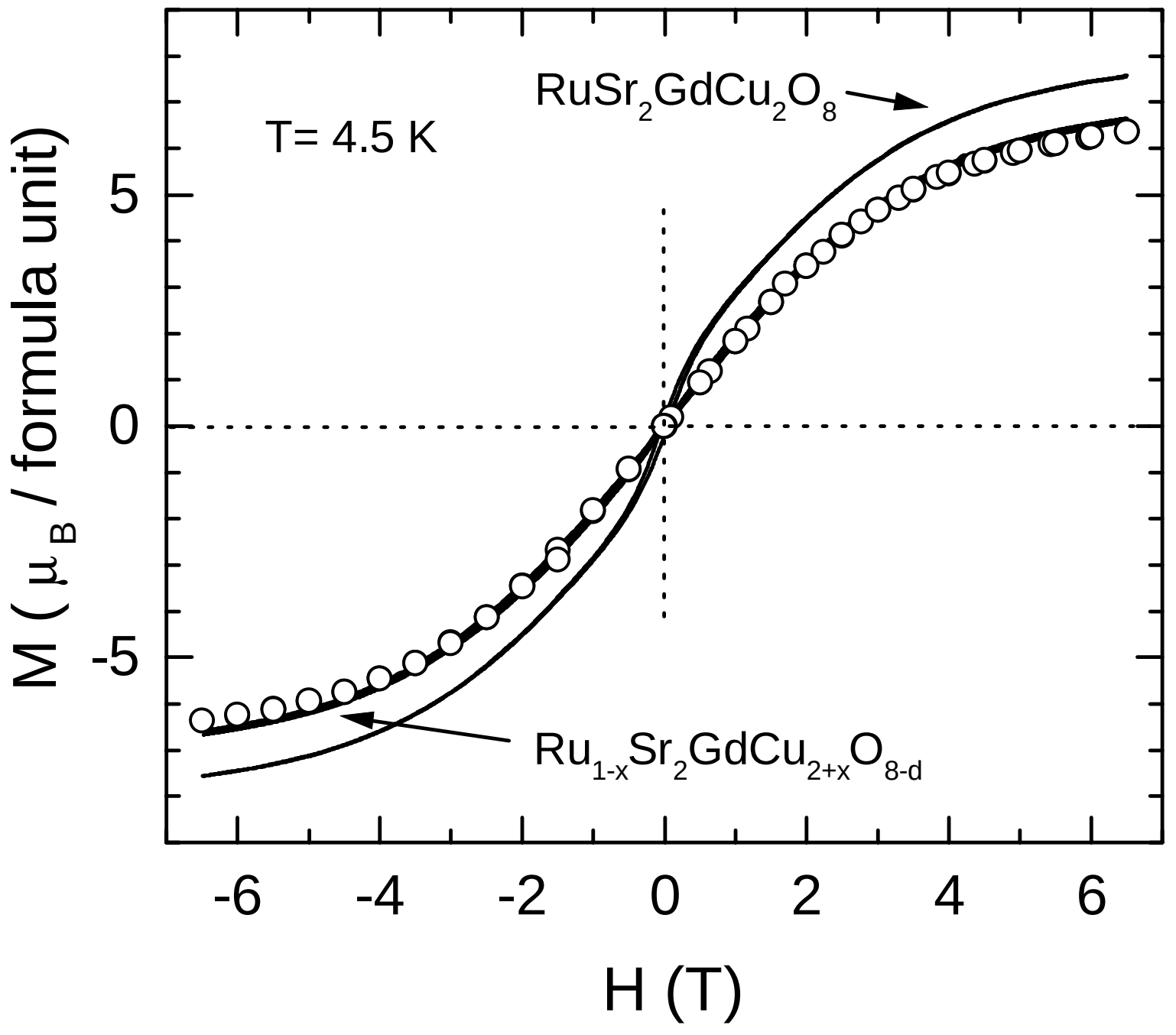

Fig. 4. M(H) dependencies at 4.5 K for $Ru_{1-x}Sr_2GdCu_{2+x}O_{8-d}$ and $GdBa_2Cu_3O_{6.2}$

It should be noted that, although the low temperature magnetization did not reveal any extra magnetic component present, we observed a slight irreversibility in field cooled (FC) and zero field cooled (ZFC) magnetization that opens below ~120 K and 100 K, for x=0.1 and 0.2, respectively [9]. This behavior could be attributed to the response of the Cu diluted Ru sublattice. Muon spin rotation (μSR) experiments performed for the x=0.1 sample indicated that the increase of the relaxation rate observed below ~120 K should not be attributed to the bulk response of the material. It can be tentatively assumed that this ZFC-FC irreversibility, if not reflecting the magnetic response of diluted $RuO_2$ planes, arises from compositional inhomogeneity (for example, the formation of Ru rich clusters in the Ru/Cu-O planes). For the x=0.3, 0.4 and 0.75 compositions we did not observe any irreversibility of the magnetization in the normal state.

The temperature dependencies of the *ac* susceptibility (Fig.2) always show two characteristic temperatures for the onset of the superconducting transitions (see inset to this figure). The onset of the intrinsic transition at $T_{C1}$ (marked with squares) and the $T_{C2}$ (marked with crosses) at which the susceptibility changes slope in reflecting the establishment of the bulk superconducting screening currents. It is worth noting that the diamagnetic contribution to the signal measured between $T_{C1}$ and $T_{C2}$ substantially increases with x along the series.

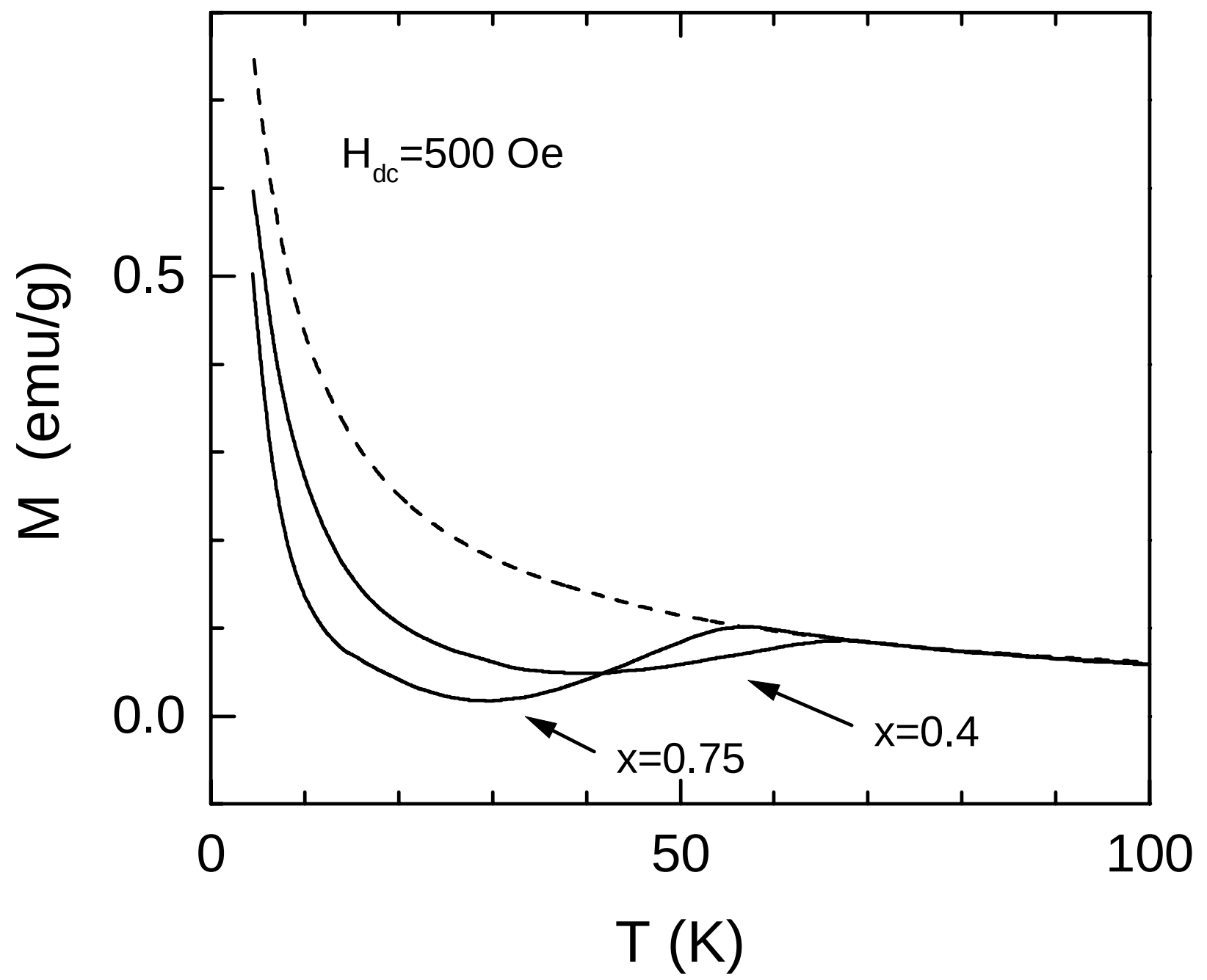

Fig. 5. *dc* magnetization for x=0.4 and x=0.75 powder samples. Dashed line: non-superconducting $GdBa_2Cu_3O_{6.2}$. $H_{dc}$= 500 Oe

Figure 5 depicts the low temperature reentrant behavior of the magnetization at $H_{dc}$=500 Oe found for every composition in the series (x=0.4 and 0.75 shown) at sufficient values of the magnetic field [9]. The results presented in this figure show the response of the samples, which were powdered to minimize the intergrain diamagnetic screening at low temperatures. The dashed line in this figure shows the behavior of $Gd^{3+}Ba_2Cu_3O_{6.2}$ that again delineates the low temperature paramagnetic contribution of the $Gd^{3+}$ ions. Regardless of this reentrant behavior of magnetization, the low temperature zero resistivity state is preserved at much higher magnetic fields (H=6.5 T was the highest field used). In [9] we propose that the paramagnetic response in the presence of superconductivity in these samples can be qualitatively understood assuming quasi-two-dimensional character of superconducting layers that are separated by non-superconducting regions. For polycrystalline samples with randomly oriented crystallites, the paramagnetic response would arise from the crystallites for which superconducting layers are oriented parallel to the external field that can penetrate the space between them. Similar effect was recently proposed to explain the anisotropy of the magnetic susceptibility (for H $\perp$ *ab* and H || *ab*) observed in highly oxygen deficient (i.e. strongly underdoped) superconducting $GdBa_2Cu_3O_{7-d}$ single crystals [10].

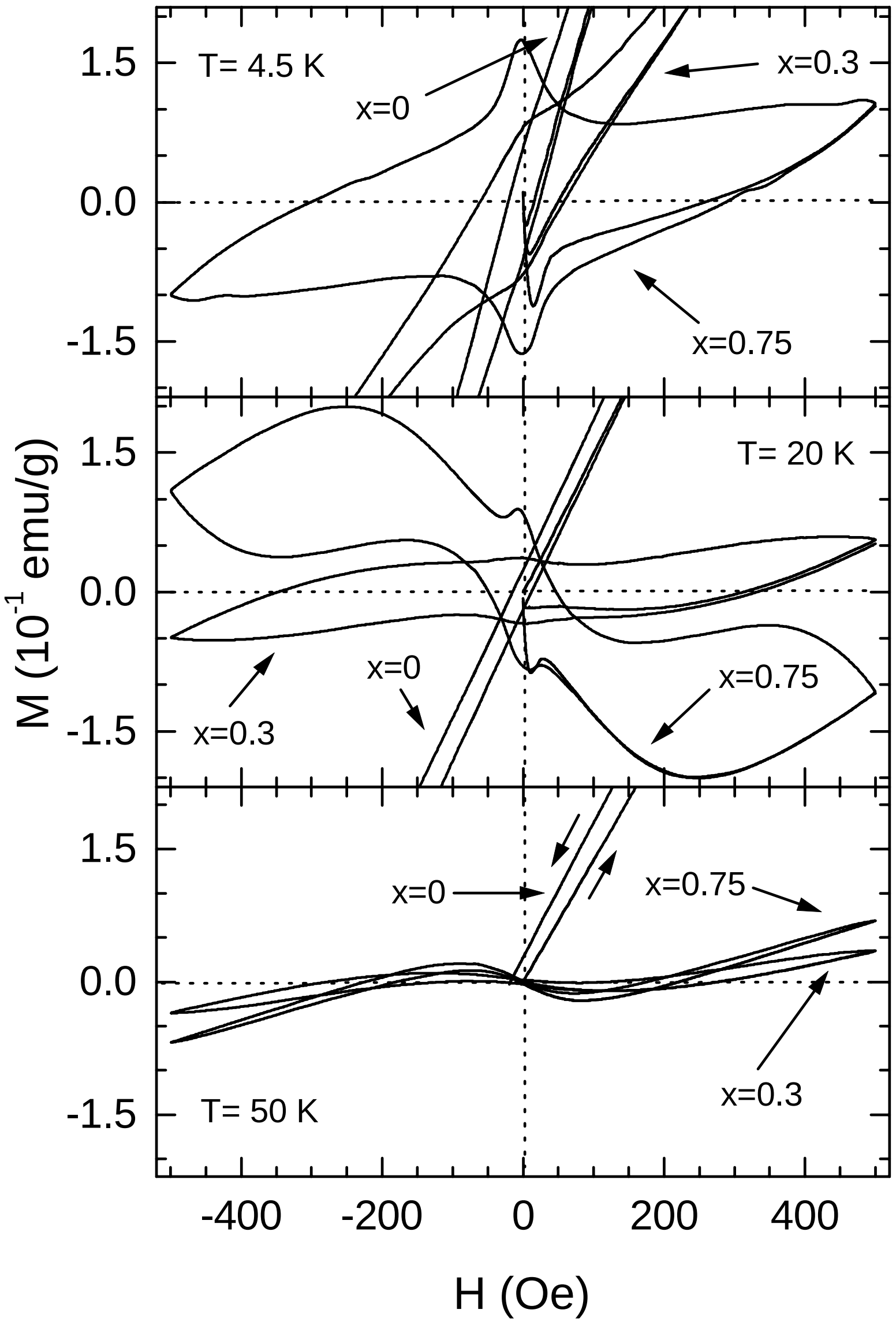

Fig. 6. The magnetic field dependencies of *dc* magnetization measured at 4.5, 20 and 50 K for x=0, 0.3 and 0.75 samples. The field cycled between -500 and 500 Oe.

Fig. 6 presents the M(H) dependencies measured for the x= 0, 0.3 and 0.75 samples at 4.5, 20, and 50 K and small magnetic fields from -500 Oe to 500 Oe. The hysteresis loops can be interpreted as the superposition of the magnetic and superconducting components. The first penetration fields at T=4.5 K, see the local minima at corresponding virgin parts of M(H) loops, are 5, 8 and 14 Oe for x=0, 0.3 and 0.75, respectively. Larger positive contributions to the magnetization are observed for smaller x (see also Fig.5). This can suggest the more constrained dimensionality of the superconducting phase for samples with smaller x, in particular for the parent $RuSr_2GdCu_2O_8$. For this compound, the additional magnetic component originating from the weak-ferromagnetism of the Ru sublattice also contributes to the magnetization above $T_C$ and below $T_N \approx 130$ K. With regard to the data presented in figure 6, we should note that for $RuSr_2GdCu_2O_8$ the ferromagnetic coercive field at 4.5 K is approx. 400 Oe when measured after the magnetic field was cycled up to the fully reversible range of magnetization. Contrary to the behavior of this compound, for all x≠0 samples the remnant magnetization was observed only at the temperatures below the superconducting transition and thus can be attributed exclusively to the irreversible movement of the vortices in the material.

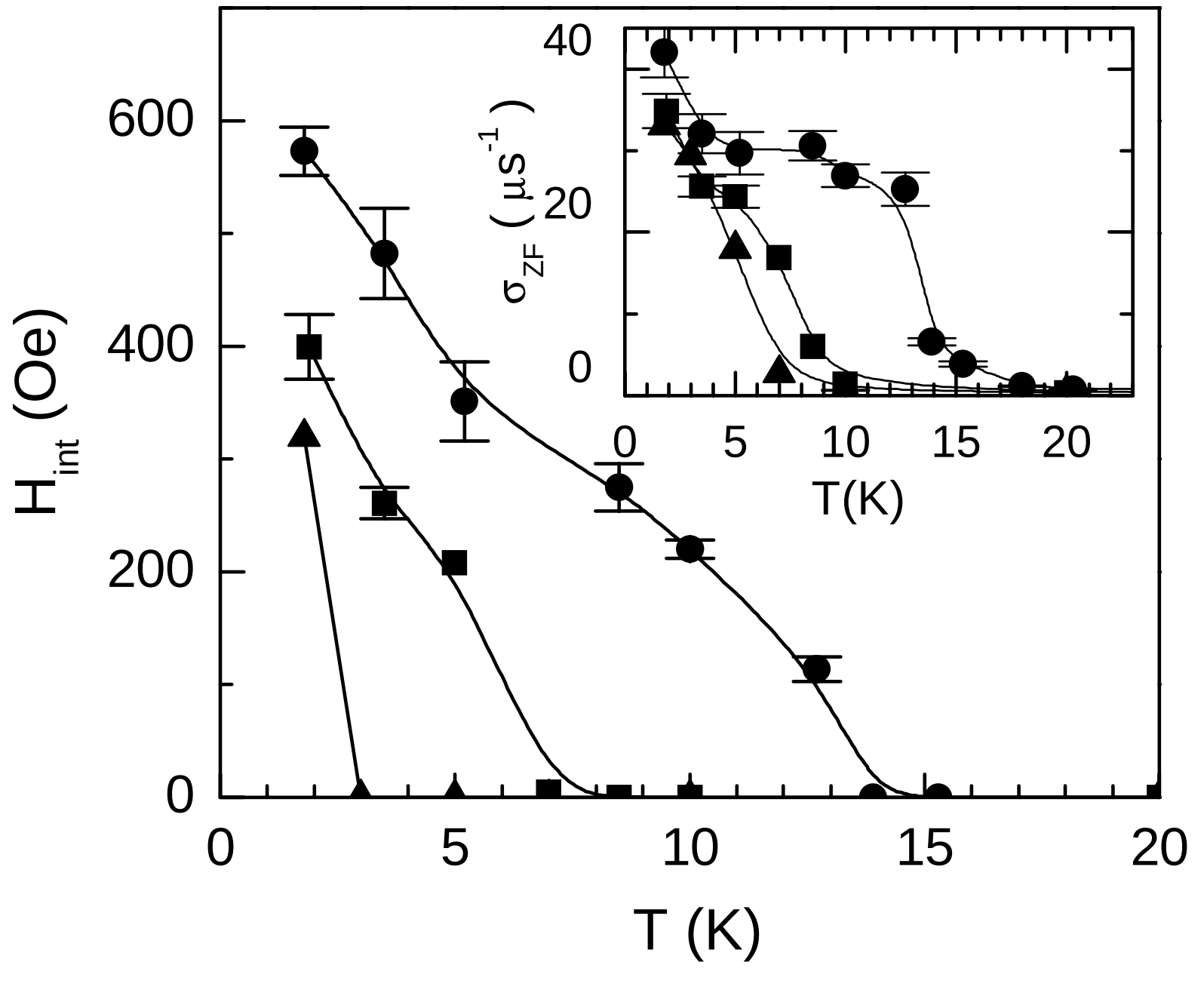

Fig. 7. Temperature dependencies of the internal magnetic field for $Ru_{1-x}Sr_2GdCu_{2+x}O_{8-d}$ measured in zero-field muon spin rotation experiment. Samples: x=0.1 (circles), 0.3 ( squares), and 0.4 (triangles). Inset shows the corresponding temperature dependencies of the relaxation rate σ.

In order to further investigate the low temperature magnetic behavior of $Ru_{1-x}Sr_2GdCu_{2+x}O_{8-d}$ we performed a series of temperature dependent muon spin rotation (μSR) experiments. This method is especially suitable to revealing the presence of even weak magnetic correlations persisting in the superconducting state in the samples [11]. Fig. 7 presents the results of zero-field muon spin rotation measurements for x=0.1, 0.3 and 0.4 compositions. The initial asymmetry parameters measured in this experiment allow us to conclude that we observe bulk (90-100% of sample volume) magnetic transitions resulting in the presence of an internal magnetic field below $T_m$=13, 6 and 2 K, for x=0.1, 0.3 and 0.4 respectively. Values of the internal magnetic field, as well as the relaxation rates presented in this figure, have been calculated from the time dependent spectra fit with an equation:

$$A = \frac{2}{3} B\cos(\gamma_\mu B_\mu t + \phi)$$

for the time evolution of the muon spin polarization described by:

$$P(t) = A\exp(-\frac{1}{2}(\gamma_\mu \Delta B_\mu t)^2) + \frac{1}{3} B\exp(-\lambda t),$$

where B is the initial asymmetry measured in the external field in the normal state,
$\gamma_\mu$=851.4 MHz/T, and $B_\mu$ is the average internal field at the muon site.

Since the sublattice of paramagnetic $Gd^{3+}$ moments orders antiferromagnetically at 2.8 K [4, 12], the Ru/Cu-diluted sublattice or Cu moments in the $CuO_2$ planes seem only to be candidate systems that can be responsible for observed behavior. Thus, assuming the bulk nature of the superconducting phase, the μSR data should be interpreted as indicative of a coexistence of the AF and SC order parameter as has been observed in other underdoped systems: see Ref. 11 for the AF correlations in underdoped $La_{2-x}Sr_xCuO_4$ and $Y_{1-x}Ca_xBa_2Cu_3O_6$ for which the microscopic inhomogeneity of the charge distribution in the $CuO_2$ planes and associated stripes formation have been proposed [13]. Further experiments aimed at understanding the microscopic nature of the low temperature magnetism observed for these compounds are in progress.

The charge carrier density at 293 K for the $Ru_{1-x}Sr_2GdCu_{2+x}O_{8-d}$ series, as estimated from Hall effect measurements ($n_H=1/[R_H \cdot e \cdot c]$) ranges from $1.7\cdot10^{27}$ (x=0) to $3.9\cdot10^{27}$ $1/m^3$ (x=0.4) and saturates for far x. Figure 8 shows the values of room temperature thermopower for the x=0, 0.1, 0.3, 0.4 samples in this series. The inset from this figure presents the temperature dependence of thermopower measured for the x=0.4 sample. Positive values in the normal state suggest the underdoped nature of this superconductor. Figure 9 presents the values of characteristic Cu-K edge energy obtained in XANES measurements for the same set of samples. All three experiments reveal that the Cu doping into the Ru positions result in enhanced effective hole doping with its probable saturation occurring for x≈0.4. This composition has the highest $T_C$ in the series (onset at 72 K vs. 45 K for x=0 and 62 K for x=0.75 [9]). We should note, that since all samples were prepared at the same oxidizing conditions, one can expect the oxygen content per formula unit could decrease with x and become closer to 7 for compositions with larger x. This effect, through its probable influence on the effective charge doping achieved, could contribute to the underdoped characteristic of the x=0.4 sample. Samples with x≠0 support variable oxygen stoichiometry that effect $T_C$, resembling the properties of Gd123. For x=0.4 composition, the post annealing at 800°C in flowing air and in 1% of oxygen decreases $T_C^{on}$ from 72 to 55 and 43 K respectively, whereas the annealing in argon at 800°C leads to the non-superconducting material. Measurements of the temperature dependencies of *dc* magnetization for high-pressure oxygen synthesized series, performed at external pressures up to 1.2 GPa reveal that $T_C$ increases at a rate of approx. 5.5 K/GPa for all investigated samples. This also suggests the underdoped character of the compounds. Fig. 10 presents the specific heat jump at the temperature of superconducting transition for $Ru_{0.6}Sr_2GdCu_{2.4}O_{8-d}$. The magnitude of $\Delta C_p/T$ is approximately 12 mJ/mol·K$^2$.

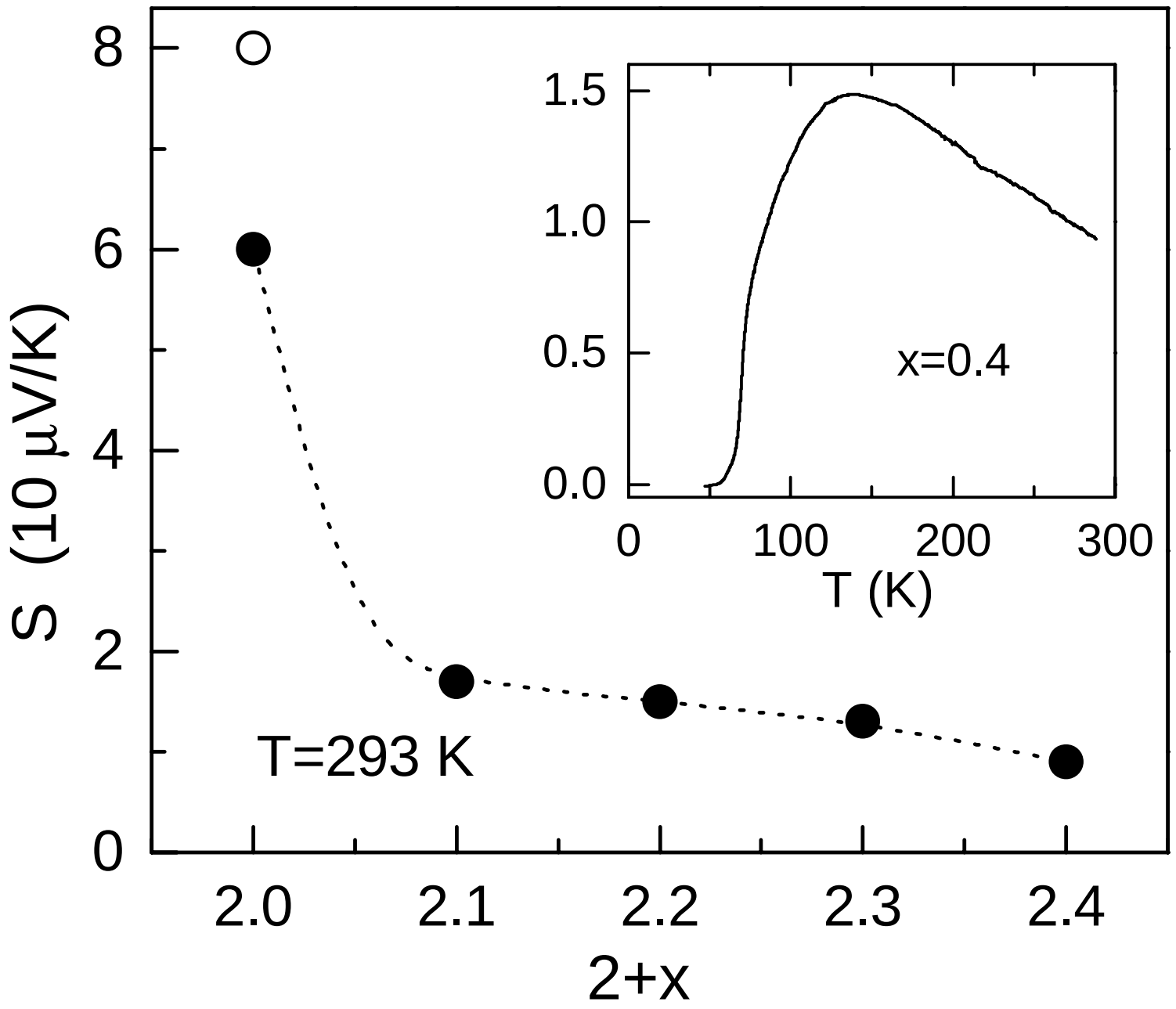

Fig. 8. Thermopower for $Ru_{1-x}Sr_2GdCu_{2+x}O_{8-d}$ superconductors. Inset shows its temperature dependence for x=0.4 sample. Open circle represents the value for non-superconducting $RuSr_2GdCu_2O_8$. T=293 K.

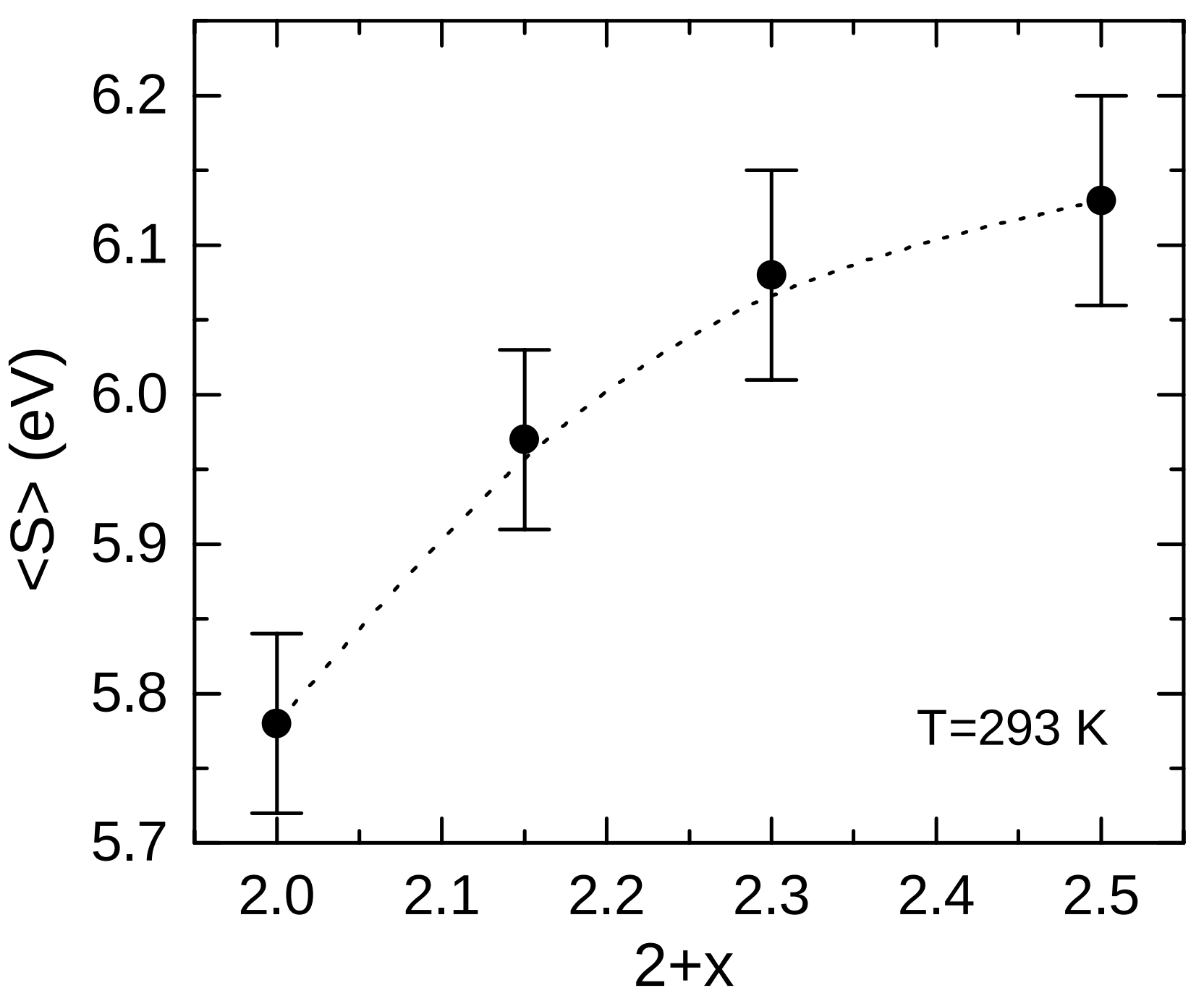

Fig. 9. XANES Cu-K edge energy vs. Cu content for $Ru_{1-x}Sr_2GdCu_{2+x}O_{8-d}$ series. T=293 K.

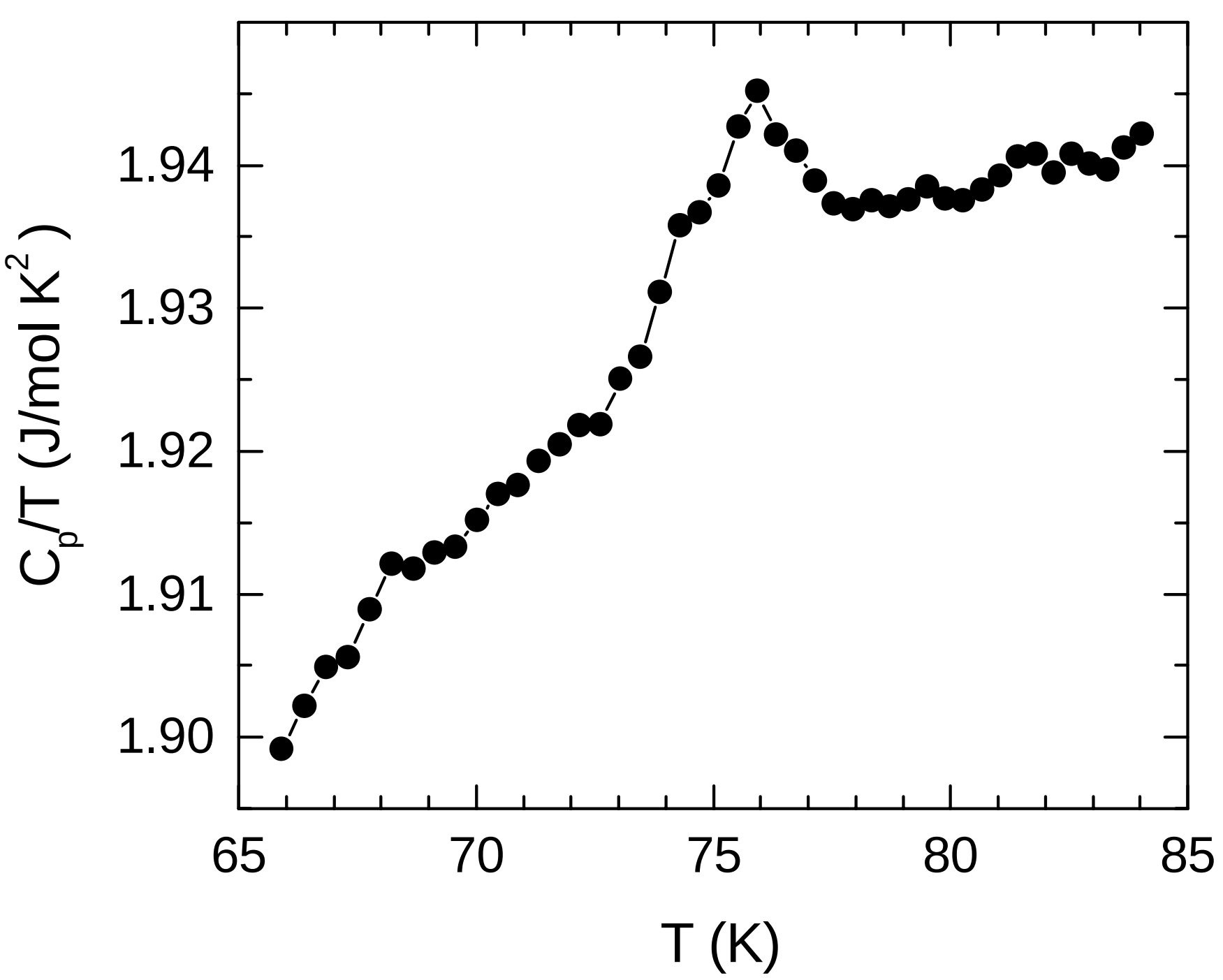

Fig. 10. $C_p/T$ vs. temperature for $Ru_{0.6}Sr_2GdCu_{2.4}O_{8-d}$ in the vicinity of superconducting transition.

Following the same high-pressure oxygen synthesis route as applied for $Ru_{1-x}Sr_2GdCu_{2+x}O_{8-d}$ materials, we have synthesized the isostructural superconducting samples of $Ru_{1-x}Sr_2EuCu_{2+x}O_{8-d}$ for x=0.4 and 0.6 ($T_C$=70 and 52 K respectively, being the onsets of resistive transitions). Since $Eu^{3+}$ ions do not carry the net magnetic moment in their ground state, these samples presented as with an opportunity to investigate their low temperature properties with no extra paramagnetic contribution present as with Gd based compounds.

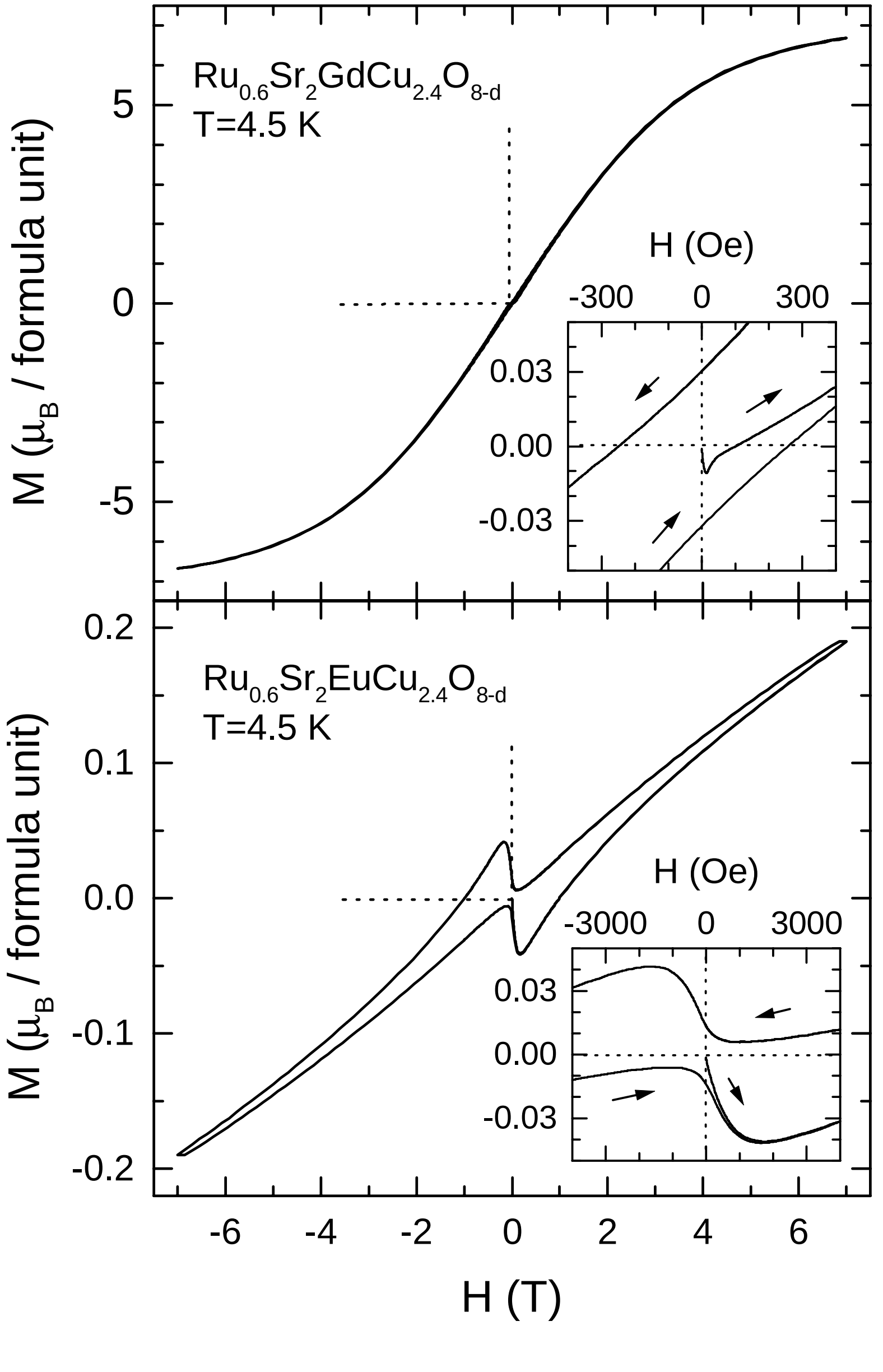

Fig. 11. M(H) dependencies for $Ru_{0.6}Sr_2GdCu_{2.4}O_{8-d}$ and $Ru_{0.6}Sr_2EuCu_{2.4}O_{8-d}$ at 4.5 K.

Fig. 11 presents comparison of M(H) dependencies for $Ru_{0.6}Sr_2GdCu_{2.4}O_{8-d}$ and $Ru_{0.6}Sr_2EuCu_{2.4}O_{8-d}$ samples measured at 4.5 K. The comparison shows again that the positive values of magnetization measured for $Ru_{0.6}Sr_2GdCu_{2.4}O_{8-d}$ in the superconducting state (at magnetic field well below $H_{c2}$) originate from the paramagnetic response of Gd ions. A similar effect, although considerably smaller because of the negligible contribution of the Eu moments, is also observed for $Ru_{0.6}Sr_2EuCu_{2.4}O_{8-d}$ and suggests the constrained dimensionality of the superconducting phase in this compound. The ZF-μSR experiment revealed the presence of low temperature magnetism for superconducting $Ru_{0.6}Sr_2EuCu_{2.4}O_{8-d}$, which is similar to the behavior found for $Ru_{0.6}Sr_2GdCu_{2.4}O_{8-d}$ [14]. The asymmetry of the muon decay spectrum registered in a zero external field experiment at 1.8 K for $Ru_{0.6}Sr_2EuCu_{2.4}O_{8-d}$ (not shown) is indicative of the presence of fast damped oscillations due to the precession of muons' spin in an inhomogeneous internal magnetic field. The initial value of the observed asymmetry also indicated the bulk nature of the phenomena. The temperature dependence of the muon spin relaxation rate (σ; measures the rate of muon spin depolarization due to the field distribution in the specimen) measured in a transverse field experiment (TF-μSR) is presented in fig. 12. In this experiment the sample was cooled down in an external field of 2000 Oe with an orientation perpendicular to the spins of the incoming muons. The initial increase of σ at 60 K, due to the distribution of the magnetic field penetrating the superconducting volume in the form of vortices, is followed by a steep increase observed at low temperatures that can be attributed to internal magnetism in the sample. The T=0 value of σ inferred for the superconducting phase by cutting off the magnetic contribution, equals 1.6 $\mu s^{-1}$ and scales well with $T_C$ according to the universal Uemura relation for underdoped HTSC compounds. The inset from figure 12 presents the temperature dependence of the internal magnetic field as calculated from the frequency of asymmetry oscillations at different temperatures. The field decreases at the superconducting transition (Meissner effect) and then increases at low temperatures in the magnetically ordered state that sets in below $T_m \approx 5$ K. The low temperature magnetism detected for $Ru_{0.6}Sr_2EuCu_{2.4}O_{8-d}$ should also be considered for its possible positive contribution to the field dependent low temperature magnetization of this specimen (Fig.11).

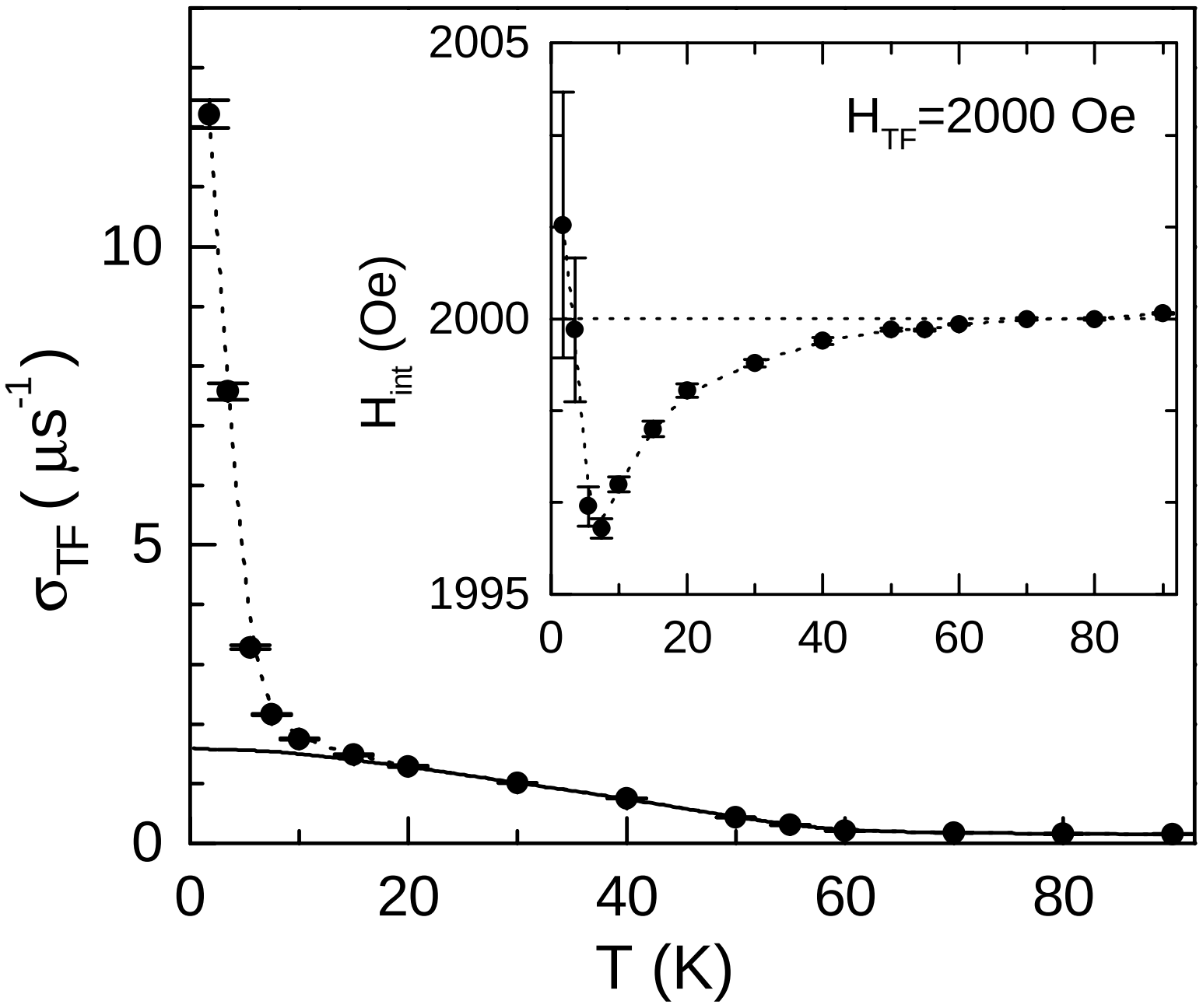

Fig. 12. Temperature dependence of the muon spin relaxation rate σ calculated from transverse field ($H_{TF}$=2000 Oe) μSR spectra for $Ru_{0.4}Sr_2EuCu_{2.6}O_{8-d}$ sample. Inset shows temperature dependence of the internal magnetic field.

**2. $RuSr_2Gd_{1-y}Ce_yCu_2O_8$ ($0 \le y \le 0.1$)**

In a separate approach to address the question of which electronic and structural parameters control $T_C$ and $T_N$ in $RuSr_2GdCu_2O_8$, we partially substituted trivalent Gd by $Ce^{4+}$ ions. This substitution by intervening into the layer located between the two $CuO_2$ planes, resembles substitutions frequently studied in other cuprates where the hole doping, and thus $T_C$, can be effectively controlled by varying the amount of substituted ion. By leaving the Ru and Cu sites intact, we attempted to achieve the modification of magnetic and superconducting properties of $RuO_2$ and $CuO_2$ layers primarily as a function of charge doping.

Polycrystalline samples of $RuSr_2Gd_{1-x}Ce_xCu_2O_8$ (x = 0, 0.02, 0.05, 0.1) were synthesized by solid state reaction of stoichiometric oxides of $RuO_2$, $CeO_2$, $Gd_2O_3$, CuO and $SrCO_3$. After calcination in air at 900°C, the material was ground, pressed into pellets and annealed in flowing Ar at 1010 °C. This minimizes the preformation of the $SrRuO_3$ impurity phase. Subsequently, the samples were annealed in flowing oxygen at increasing temperatures from 1030 °C to 1060 °C with frequent intermediate grinding and pelletizing. The resulting material was slowly cooled to 500 °C and then quenched to room temperature. The structure of all samples was indexed in the tetragonal 4/mmm symmetry similar to the $Ru_{1-x}Sr_2GdCu_{2+x}O_{8-d}$ compounds. We found the lattice constants increase with y [15], which suggests effective electron doping into the antibonding part of Cu-O orbitals. This resembles the effect observed in superconducting $Nd_{2-x}Ce_xCuO_{4\pm d}$ a-axis dimensions.

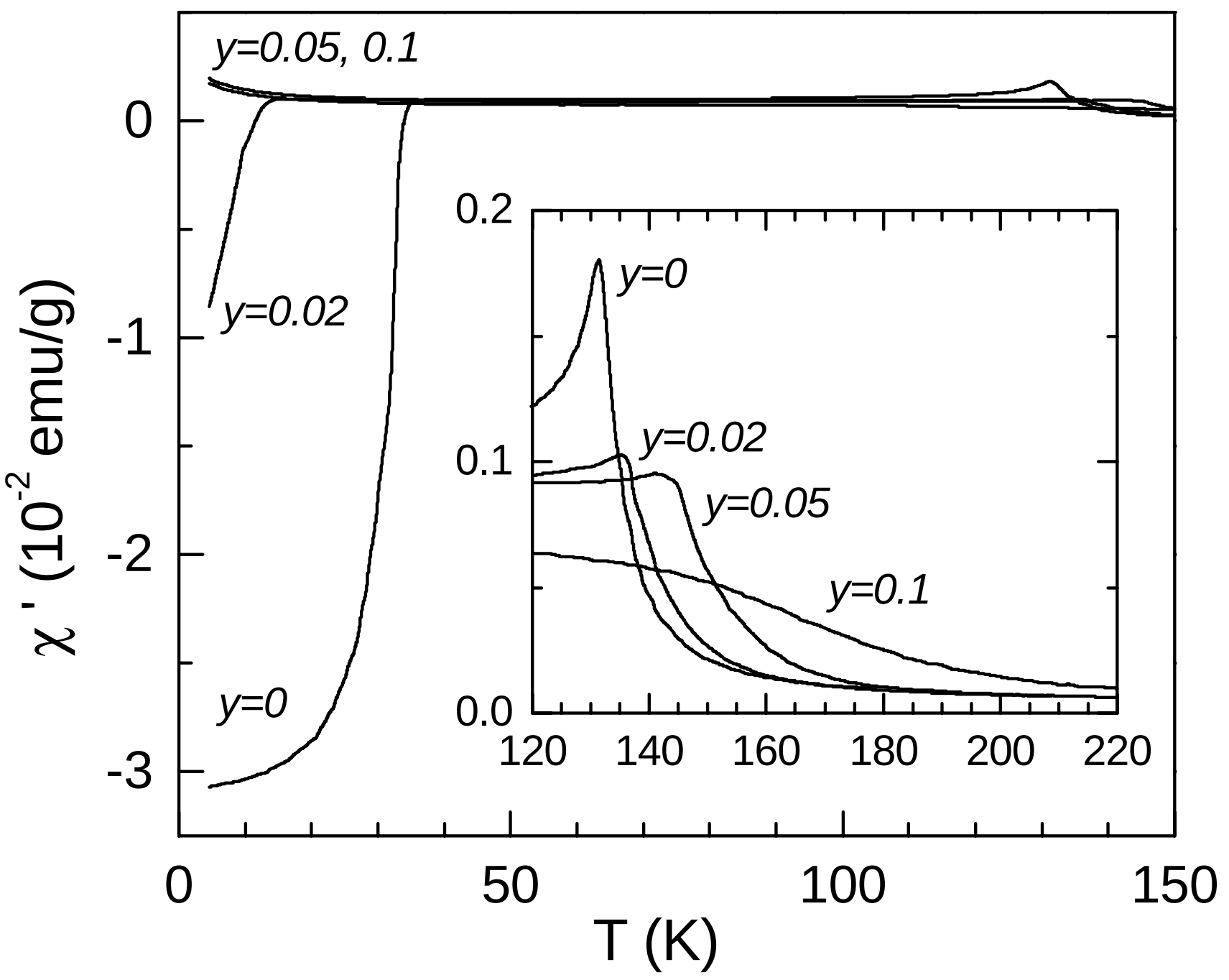

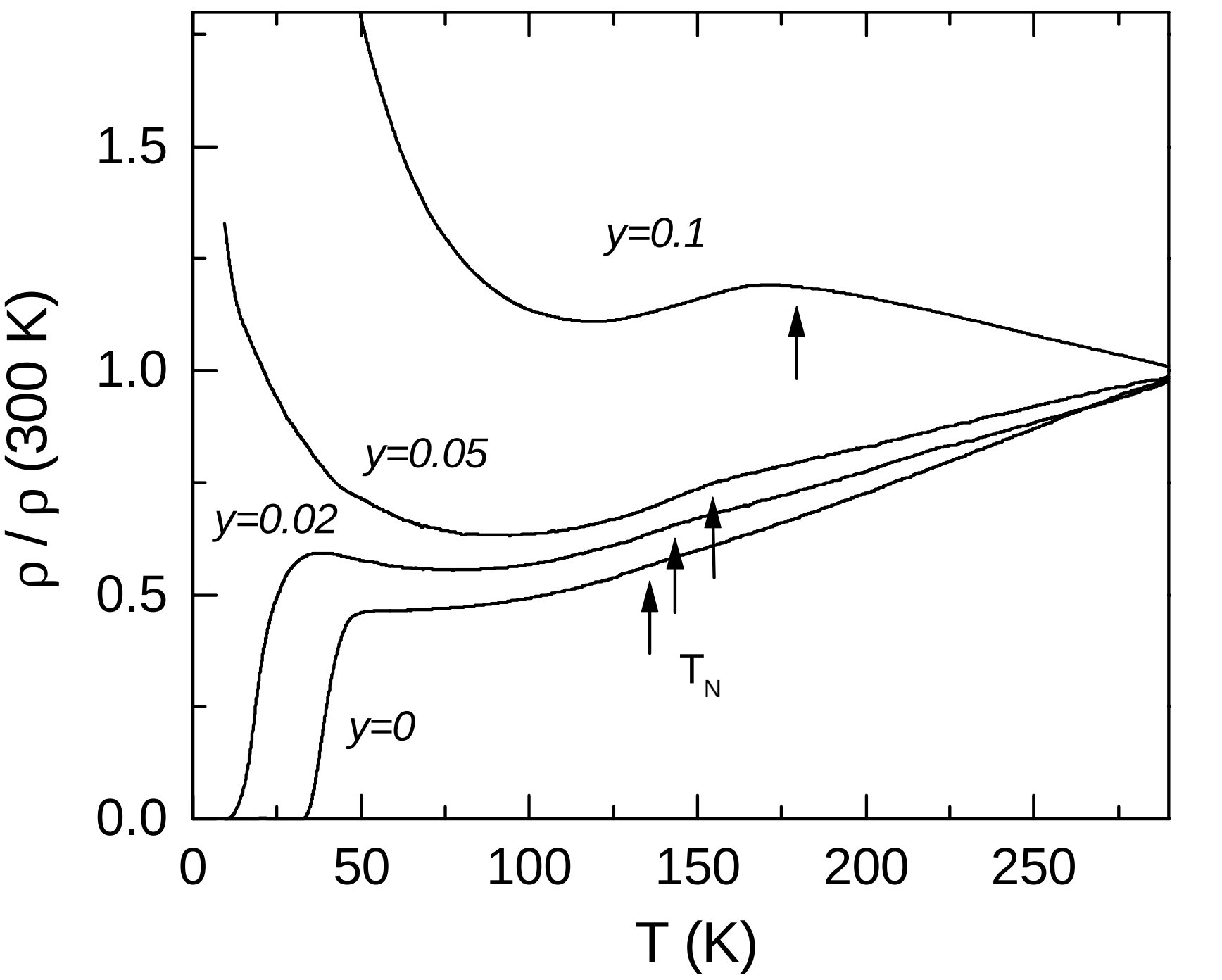

Figs. 13 and 14. The temperature dependencies *ac* susceptibility and resistivity for $RuSr_2Gd_{1-y}Ce_yCu_2O_8$.

Figs. 13 and 14 present the temperature dependencies of the *ac* susceptibility and resistivity for this series. For y=0.02, $T_C^{on}$ decreases to 36 K and an upturn of the resistivity above this transition is indicative of increased charge localization. The y=0.05 and 0.1 samples are not superconducting. The slight decrease of the resistivity around 130 K for x=0 corresponds to the temperature of magnetic ordering of the Ru sublattice. This agrees with a similar feature reported previously [2]. The decrease of resistivity at $T_N$ becomes more pronounced with increasing Ce substitution. This can be interpreted as increased contribution from $RuO_2$ layers, in their magnetically ordered state, to the conductivity of the system; or alternatively as the indication of charge redistribution between $RuO_2$ and $CuO_2$ layers which takes place at $T_N$.

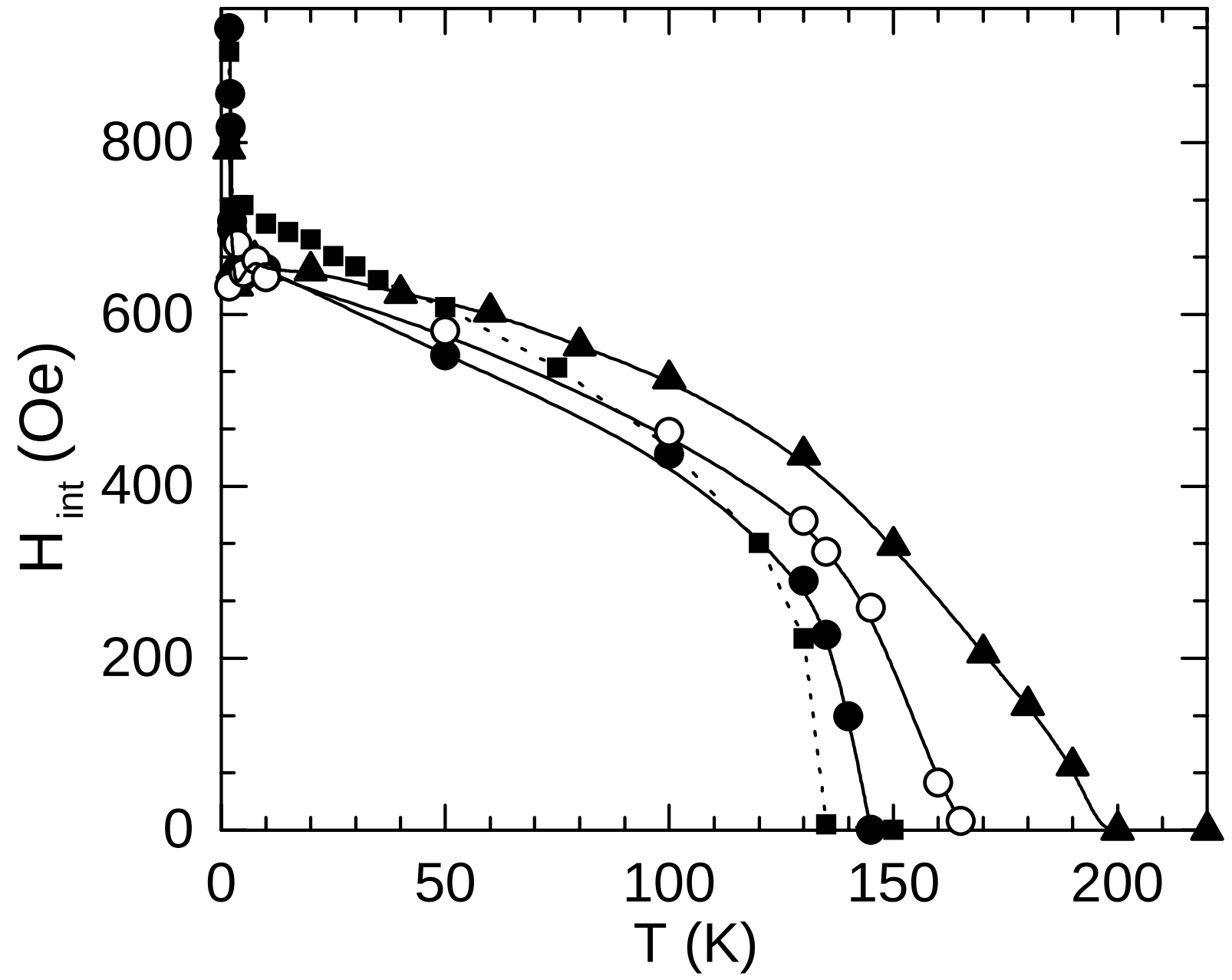

Fig. 15. Temperature dependencies of the internal magnetic field for $RuSr_2Gd_{0.9}Ce_{0.1}Cu_2O_8$ (triangles) and non-superconducting samples of $RuSr_2GdCu_2O_8$ (closed circles) and $RuSr_2EuCu_2O_8$ (open circles) measured in zero-field μSR experiment. Closed squares show data for superconducting $RuSr_2GdCu_2O_8$ after Ref.2.

The onset of the superconducting transition as seen in the real part of *ac* susceptibility (χ') (Fig. 13) coincides with the temperatures at which the material attains zero resistivity (Fig.14). The maxima of χ' near and above 130 K distinguish the temperatures of the magnetic ordering of the Ru sublattice ($T_N$). The $T_N$, as also reflected on the resistivity dependencies, increases with Ce doping. Fig. 15 (triangles) presents the temperature dependence of the internal field for $RuSr_2Gd_{0.9}Ce_{0.1}Cu_2O_8$, calculated from the results of zero-field μSR measurement. The temperature of magnetic ordering of the Ru sublattice can be estimated to $T_N$≈195 K; which is in quite good agreement with magnetic and transport data. Fig. 15 also presents temperature dependencies of the internal field measured for $RuSr_2GdCu_2O_8$ and $RuSr_2EuCu_2O_8$ – these will be addressed further in a later part of the article.

Because the observed decrease of $T_C$, as well as changes of the lattice parameters with y [15] remains consistent with the expected effect of electron doping achieved by heterovalent Ce substitution, we can attempt to combine the characteristics of $RuSr_2Gd_{1-y}Ce_yCu_2O_8$ and $Ru_{1-x}Sr_2GdCu_{2+x}O_{8-d}$ compounds to construct a qualitative phase diagram that presents properties of both series vs. changing hole doping. Fig. 16 illustrates this approach, where the horizontal axis reflects the Ce→Gd and Cu→Ru doping, for left and right parts of the diagram respectively. The important general conclusion is how the superconducting and magnetic properties depend on the charge doping in the Ru-1212 system, and, quite unexpectedly, that the low temperature magnetic correlations seen in μSR experiments seem to be well preserved into the superconducting state. That of course raises the questions regarding the microscopic nature of this magnetic ordering, and description for its' coexistence with superconductivity. Both questions expose the challenge that currently defines our research effort.

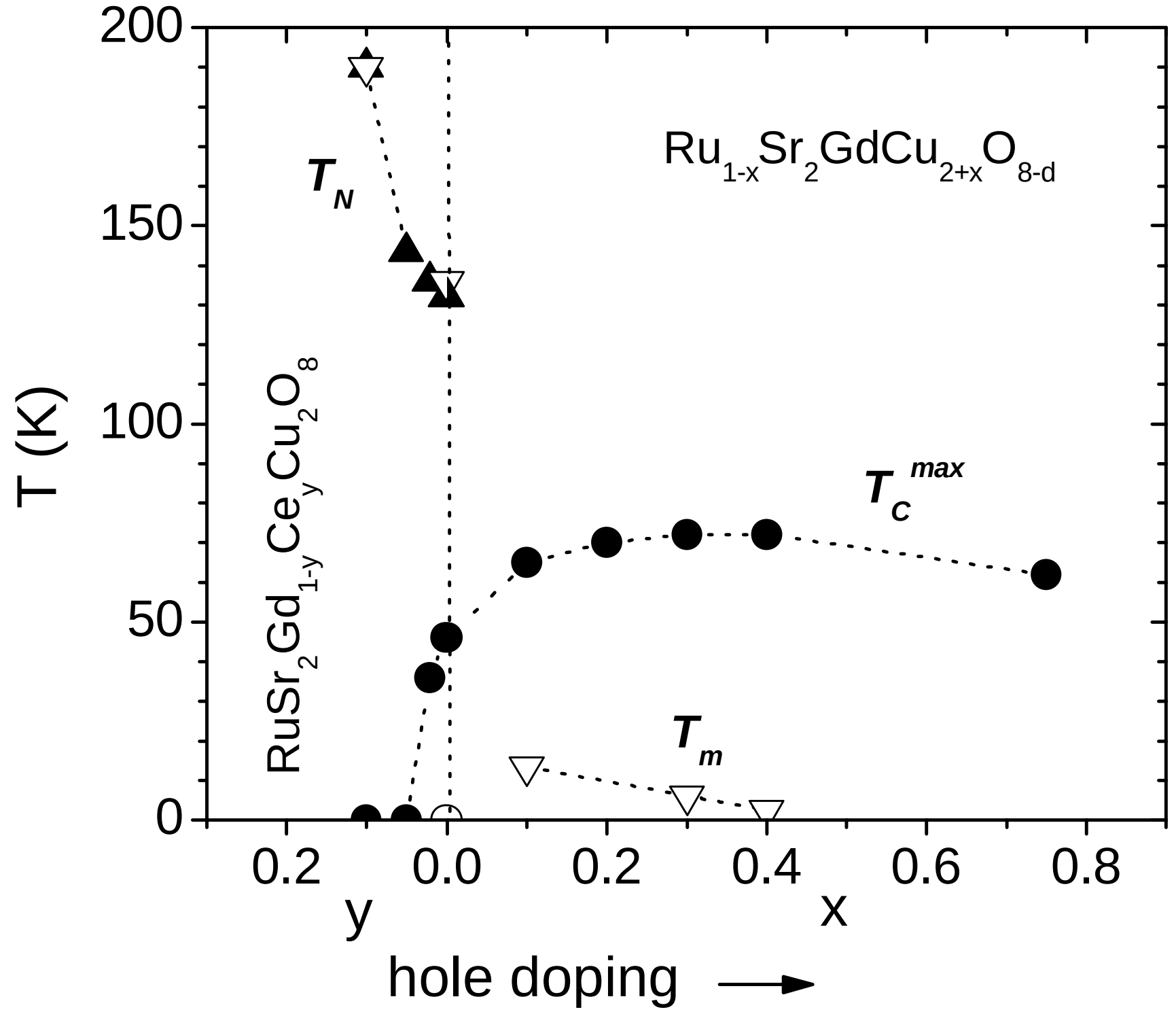

Fig. 16. Properties of $RuSr_2Gd_{1-y}Ce_yCu_2O_8$ and $Ru_{1-x}Sr_2GdCu_{2+x}O_{8-d}$ vs. Ce→Gd and Cu→Ru substitutions, which scales with hole doping in the series. Open triangles: temperatures of magnetic phase transitions ($T_N$, $T_m$), as determined from temperature dependencies of the internal field measured in zero-field μSR experiment. Closed triangles and circles: temperatures of the magnetic ($T_N$) and superconducting ($T_C$) phase transitions, as from the results of magnetic and transport measurements.

### 3. Superconducting and non-superconducting samples of $RuSr_2RECu_2O_8$ (RE=Gd, Eu)

The usual route for the synthesis of 1212-type ruthenocuprates ([2, 3] or synthesis of $RuSr_2Gd_{1-y}Ce_yCu_2O_8$ as presented above) results in superconducting material with maximum $T_C$=45 K for RE=Gd, and a somewhat lower value of $T_C$ reported for Eu. Recently, we have found [14] that modified synthesis conditions, with a final annealing at 930°C in 1% of oxygen in argon, lead to non-superconducting single-phase samples for both Gd and Eu based Ru-1212. Fig.17 presents the temperature dependence of *dc* magnetization for non-superconducting $RuSr_2GdCu_2O_8$. The inset (1) shows the low temperature magnetization, which, as one could expect, forms a maximum at 2.8 K – the temperature of the antiferromagnetic phase transition of the Gd sublattice.

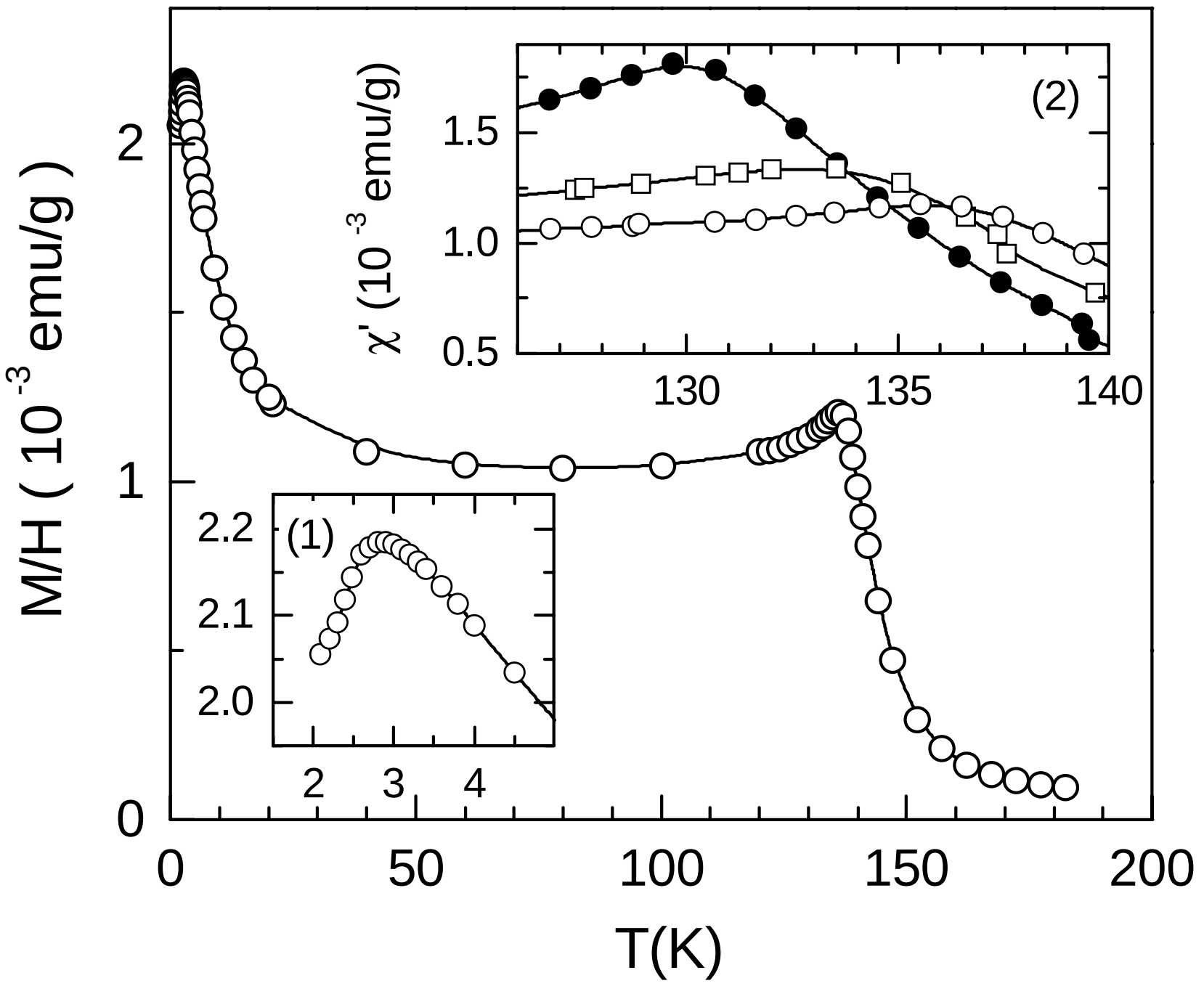

Fig. 17. Temperature dependence of *dc* susceptibility for non-superconducting $RuSr_2GdCu_2O_8$. Inset (1) – low temperature behavior on the expanded scale. Inset (2) - *ac* susceptibility ($H_{ac}$=1 Oe, f=200 Hz) of $RuSr_2GdCu_2O_8$ at the temperature range of magnetic phase transition in Ru sublattice; open circles: for non-superconducting material synthesized at 930°C in 1% of oxygen in argon, open squares: for the same sample after additional 24 hour annealing at 1060°C in oxygen ($T_C^{on}$≈5 K), closed circles: for the same sample after additional 180 hour annealing at 1060°C in oxygen that led to an inducement of superconductivity with $T_C^{on}$≈45 K.

Interestingly, our thermogravimetric measurements did not show any noticeable changes in the oxygen content between superconducting and non-superconducting materials. Thus, we should consider that the structural differences between both compounds occur in the form of slight, but still decisive, changes of cation site defects/substitutions or in the difference in the structural distortions. Comparison of the magnetic properties of superconducting and non-superconducting samples of $RuSr_2GdCu_2O_8$, utilizing the *dc* magnetization, *ac* susceptibility [14] and μSR data (see Fig. 15), reveal that the weak ferromagnetic component is always enhanced for the superconducting compounds, whereas the temperature of the magnetic ordering is slightly lowered (130 vs. 136 K). The hole doping, as indicated by the values of thermopower presented in Fig.8, is enhanced for the superconducting compound (compare open and closed circles for x=0). Superconducting $RuSr_2EuCu_2O_8$ can still be obtained by repeated oxygen annealing at 1060°C, however, our experiments indicate that this occurs only for materials containing a small fraction of secondary phases (predominantly $SrRuO_3$) [14]. That, in turn, can create a favorable situation for the formation of structural defects. The superconductivity in these samples should be considered within the scope of our finding that partial Cu→Ru substitution in $Ru_{1-x}Sr_2RECu_{2+x}O_{8-d}$ (RE=Gd, Eu) leads to a significant increase of the superconducting $T_C$. The superconducting $RuSr_2EuCu_2O_8$ for which neutron diffraction results are reported in [16] has an even smaller magnetic transition temperature ($T_N$≈120 K**)** than its Gd based superconducting counterpart. ZF-μSR data reveal that for the non-superconducting single phased $RuSr_2EuCu_2O_8$ the magnetic order exists below approximately 150 K (see Fig. 15). Since the *ac* susceptibility of $RuSr_2EuCu_2O_8$ below $T_N$ remains significantly smaller than for $RuSr_2GdCu_2O_8$, and the paramagnetic contribution of the $Gd^{3+}$ sublattice is too small to fully account for this difference (see our results presented in [14]), one can conclude that the weak-ferromagnetic component in $RuSr_2GdCu_2O_8$ is always enhanced compared to its Eu-based analogue compound. Detailed neutron diffraction experiments on $Gd^{160}$ enriched Ru-1212, to further elucidate on the role of structural disorder in determining the properties of this material are currently in progress.

## 4. Conclusions

We reported two heterovalent substitutions in $RuSr_2RECu_2O_8$ that expand this parent Ru-1212 ruthenocuprate to two new series of compounds: hole doped $Ru_{1-x}Sr_2GdCu_{2+x}O_{8-d}$ and electron doped $RuSr_2Gd_{1-y}Ce_yCu_2O_8$. The characteristics of these materials allow us to propose the qualitative phase diagram that links their properties to different hole doping realized in the series. The magnetic properties of the series of $Ru_{1-x}Sr_2GdCu_{2+x}O_{8-y}$ superconductors (maximum $T_C$=72 K for x=0.4) reveal the dominant contribution of the paramagnetic response of the $Gd^{3+}$ sublattice at low temperatures. The results are interpreted as indicative of the constrained dimensionality of the superconducting phase that apparently evolve along the series toward the quasi-two dimensional behavior characteristic for the x=0 parent compound.

The non-superconducting samples of $RuSr_2RECu_2O_8$ (RE=Gd, Eu) were synthesized by modified synthesis at 930C in 1% of oxygen in argon. Comparison of the magnetic properties between superconducting and non-superconducting compounds of the same nominal composition reveal that the weak-ferromagnetic component is always enhanced for the superconducting material, whereas the temperature of the magnetic ordering is always slightly lowered. The effect can be attributed to different levels of cation site defects/substitutions or to the difference in details of structural distortions present in the crystal structure. This, in turn, would influence the weak ferromagnetic response observed for the antiferromagneticaly ordered and canted sublattice of the Ru moments. We also should consider that even minute effective Cu→Ru substitution can provide the hole doping mechanism that stabilizes the superconducting phase in the system.

**Acknowledgement**

Research was supported by the National Science Foundation (DMR-0105398), and by the State of Illinois under HECA. P.W.K., A.S., R.K., I.S. and B.D. would like to thank Dr. D. Herlach and Dr. A. Amato of PSI, Villigen for their valuable assistance with the µSR experiments. S.M.M. acknowledges support by the NSF (CHE-9871246 and CHE-9522232) and the use of the Advanced Photon Source was supported by the U.S. Department of Energy, Basic Energy Sciences, Office of Science, under Contract No. W-31-109-Eng-38. A.W., R.P., and I.F acknowledge the Polish State Committee KBN contract No. 5 P03B 12421.